\documentclass[10pt,conference]{IEEEtran}

\usepackage{array}
\newcolumntype{L}[1]{>{\raggedright\let\newline\\\arraybackslash\hspace{0pt}}m{#1}}
\newcolumntype{C}[1]{>{\centering\let\newline\\\arraybackslash\hspace{0pt}}m{#1}}
\newcolumntype{R}[1]{>{\raggedleft\let\newline\\\arraybackslash\hspace{0pt}}m{#1}}
\usepackage{makecell}
\usepackage{tabularx}
\usepackage{booktabs}

\usepackage{cite}
\usepackage{amsmath,amssymb,amsfonts}
\usepackage{algorithmic}
\usepackage{graphicx}
\usepackage{textcomp}
\usepackage{xcolor}
\def\BibTeX{{\rm B\kern-.05em{\sc i\kern-.025em b}\kern-.08em
    T\kern-.1667em\lower.7ex\hbox{E}\kern-.125emX}}
\begin{document}

\title{The Type to Take Out a Loan? A Study of Developer Personality and Technical Debt}

\author{\IEEEauthorblockN{Lorenz Graf-Vlachy\IEEEauthorrefmark{1}\IEEEauthorrefmark{2} and Stefan Wagner\IEEEauthorrefmark{1}}
\IEEEauthorblockA{
\IEEEauthorrefmark{1}\textit{Institute of Software Engineering, University of Stuttgart}\\ Stuttgart, Germany\\
\IEEEauthorrefmark{2}\textit{TU Dortmund University}\\ Dortmund, Germany}
\{lorenz.graf-vlachy\textbar stefan.wagner\}@iste.uni-stuttgart.de
}

\maketitle

\begin{abstract}
\textit{Background:} Technical debt (TD) has been widely discussed in software engineering research, and there is an emerging literature linking it to developer characteristics. However, developer personality has not yet been studied in this context. \textit{Aims and Method:} We explore the relationship between various personality traits (Five Factor Model, regulatory focus, and narcissism) of developers and the introduction and removal of TD. To this end, we complement an existing TD dataset with novel self-report personality data gathered by surveying developers, and analyze 2,145 commits from 19 developers. \textit{Results:} We find that conscientiousness, emotional stability, openness to experience, and prevention focus are negatively associated with TD. There were no significant results for extraversion, agreeableness, promotion focus, or narcissism. \textit{Conclusions:} We take our results as first evidence that developer personality has a systematic influence on the introduction and removal of TD. This has implications not only for future research, which could, for example, study the effects of personality on downstream consequences of TD like defects, but also for software engineering practitioners who may, for example, consider developer personality in staffing decisions.
\end{abstract}


\section{Introduction}

\subsection{Technical Debt and its Consequences}

Ever since Cunningham famously remarked that ``Shipping first time code is like going into debt''~\cite[p. 30]{Cunningham.1992}, the
notion of ``technical debt'' (TD) has gained traction in practitioner circles and academia alike~\cite{Ramac.2022}. Definitions of TD vary slightly, but researchers have converged on the notion that TD is the result of making technical compromises that give rise to a ``collection of design or implementation constructs that are expedient in the short term, but set up a technical context that can make future changes more costly or impossible.'' Consequently, TD ``presents an actual or contingent liability''~\cite[p. 112]{Avgeriou.2016}.

While TD can be beneficial, especially in terms of developer velocity, the literature frequently discusses its potential downsides~\cite{Besker.2017}.
One of the most frequently mentioned adverse consequences of TD is its negative impact on developer morale~\cite{Ghanbari.2017}. Developers generally dislike incurring TD~\cite{Besker.2020}, and it has been found to be psychologically taxing~\cite{Olsson.2021}. This leads to TD having demotivating effects~\cite{Ramac.2022}. Others have widened the scope and considered other consequences like reduced developer productivity due to reduced maintainability of code, decreased code quality (leading to more defects and subsequent costs), as well as increased uncertainty and risk~\cite{Tom.2013}.

\subsection{Technical Debt as Risk-Taking to be Managed}

In fact, to at least a substantial degree, the issue of TD can be interpreted as a matter of risk-taking. Prior scholars have clearly established, and it follows from the definition provided above, that TD is about making trade-off decisions~\cite{Shull.2011, Avgeriou.2016}. For one, such decisions about TD are a trade-off between saving effort in the present and having to repay this principal with interest in the future. Consequently, they constitute risk-taking, since the implied interest rate can only become fully clear in the future, which is uncertain and thus risky~\cite{Schmid.2013,Kruchten.2013b}.
For another, however, the risky nature of TD is exacerbated by the fact that not only is the interest amount unclear, but there is a ``possibility that debt may never need to be paid back''~\cite[p. 52]{Allman.2012} in the first place. Perhaps one of the most lucid discussions of this issue can be found in Schmid's work on the limits of the TD metaphor, in which he distinguishes between ``potential technical debt'', which he explicitly characterizes as being ``akin to a risk'', and ``effective technical debt''~\cite[p. 64]{Schmid.2013}, which is the amount of TD that is actually relevant to the future evolution of a software system (i.e., the part of it that will actually require future remediation efforts, but which can of course not be perfectly identified and anticipated).

Correspondingly, TD is often viewed as something that may need to be paid down at some point, but that, ad minimum, should be very consciously managed. At least two key strands of literature exist in this context.
First, researchers have spent considerable effort on measuring, that is, assessing the magnitude, of TD in a system. Researchers developed different indicators, for instance, code smells~\cite{Alves.2016}. Others developed various tools to automatically measure TD, for example using code metrics~\cite{Avgeriou.2021}. 
Several such tools, such as SonarQube, provide composite quality indicators of maintainability, which are particularly appreciated by junior developers~\cite{Gilson.2020}. Ultimately, researchers unsurprisingly made considerable efforts to estimate TD costs~\cite{Curtis.2012, Besker.2017}.

Second, researchers have concerned themselves with identifying, prioritizing, and paying down TD. Digkas \emph{et al.}, for instance, studied how technical debt is handled~\cite{Digkas.2017} and paid back~\cite{Digkas.2018} in the Apache ecosystem. Nayebi \emph{et al.} provide an extensive longitudinal case study on how architectural TD is identified and paid down in the context of a healthcare communications product. Maipradit \emph{et al.} recently developed an automated classifier to identify self-admitted TD in code~\cite{Maipradit.2020}. Going beyond such individual studies, Alves \emph{et al.} provide an overview of tools to identify and manage TD, including, for example, cost-benefit analysis and portfolio approaches~\cite{Alves.2016}.
Closely related, recent reviews of the work on the prioritization of TD (both against other TD or against new features) identified a set of strategies to prioritize and manage TD. Alfayez \emph{et al.}, for example, identified 24 different strategies~\cite{Alfayez.2020}, whereas Lenarduzzi \emph{et al.} aggregate the strategies they identified into four broader categories and one additional category of combination approaches~\cite{Lenarduzzi.2021}.

\subsection{Antecedents of Technical Debt}

In the quest to best manage TD, researchers have also explicitly begun to explore its causes and antecedents. Some researchers have, for instance, identified factors such as firms' business model innovation as a driver of TD~\cite{YliHuumo.2015}. Others found, for example, the number of commits and the number of lines of code (LOC) in a project to be positively related to TD~\cite{Bedi.2022}.
Yet another stream of research has focused on describing which incentives and punishments companies employ to motivate developers to avoid or pay down TD~\cite{Besker.2022}.
Overall, a myriad of factors appear to influence TD. Rios \emph{et al.}, for example, performed a qualitative study and identified 57 factors that drive TD, with time pressure due to looming deadlines ranking highest~\cite{Rios.2018}. In a large survey, Rios \emph{et al.} later increased that number to 78~\cite{Rios.2020}. A multi-country replication of this survey subsequently derived eight overarching categories of antecedents of TD, and again identified deadlines as the most important cause of TD~\cite{Ramac.2022}. Interestingly, the perception among practitioners regarding the key causes of TD changes with experience. More experienced software developers focus less on technical issues as causes of TD, and more on human factors~\cite{Freire.2021}.

As is thus perhaps to be expected, one specific stream of research has begun to focus on individual developers to understand the origins of TD. Amanatidis \emph{et al.}, for instance, studied various PHP projects and found that individual developers differ substantially with regard to the amounts of TD they incur~\cite{Amanatidis.2017}. Alfayez \emph{et al.} and Codabux and Dutchyn reaffirmed these findings in other codebases and went further in explicitly profiling individual developers, identifying various characteristics that are predictive of their actions regarding the introduction and removal of TD~\cite{Alfayez.2018, Codabux.2020}.

However, thus far, we are not aware of any research that links developers' personality to TD. This is surprising because it is a core tenet of psychological research that personality is strongly predictive of a wide variety of individuals' behaviors~\cite{John.2021} and one might therefore expect it to also have a bearing on whether a given developer is more or less willing to incur or pay back TD, which in turn may have important consequences, for example, for staffing project teams. Consequently, in this study, we ask the following research question: \textit{How is developer personality related to introducing and removing TD?}

\section{Background and Hypotheses}

In this paper, we study several different aspects of developer personality and their relationship with technical debt. Specifically, we consider the personality traits covered by the Five Factor Model, regulatory focus, and narcissism. Below, we lay out the theoretical background of each, provide an overview over its prior use in software engineering research, and develop hypotheses regarding how each trait might be related to changes in technical debt.

\subsection{Five Factor Model Personality Traits}

The Five Factor Model (FFM; sometimes also referred to as the ``Big Five'') is arguably the most established personality trait model in psychology~\cite{Matthews.2003, McCrae.1992, John.2021b}. It was developed using a psycholexical approach and comprises five broad traits---extraversion, agreeableness, conscientiousness, emotional stability (or its inverse, neuroticism), and openness to experience---that are rather stable across different situations. The traits are universal in that empirical evidence shows that they are equally valid for different sexes, races, cultures, and age groups~\cite{Ching.2014}.

The FFM has been used extensively in software engineering research.
The following is thus only an illustrative treatment of more recent works.\footnote{Several recent journal articles provide fairly extensive reviews of the use of the FFM in software engineering research~\cite{Iyer.2021, Calefato.2019}.}
Rastogi and Nagappan~\cite{Rastogi.2016}, for instance, studied the FFM personality traits of GitHub contributors. They found that developers who contributed more scored higher on openness to experience, conscientiousness, and extraversion, but lower on emotional stability and agreeableness. However, Calefato \emph{et al.}~\cite{Calefato.2019} later repeated this study with an improved personality measure and found no such effects.

Karimi \emph{et al.}~\cite{Karimi.2016} studied the relationship between personality and programming style and performance (including code quality) in student programmers. They found that openness to experience was positively associated with a breadth-first programming style and conscientiousness was positively associated with a depth-first style. They did not explicitly link FFM traits to programming performance, but found that a breadth-first programming style is linked to superior performance.

Paruma-Pabón \emph{et al.}~\cite{ParumaPabon.2016} captured developer FFM personality, as well as developers' needs and values, from software project mailing list emails, and clustered developers with similar personality types. They demonstrated that the personality of developers with commit privileges was linked to their behavior within the projects.

Relatedly, Calefato \emph{et al.}~\cite{Calefato.2019} studied the personality of Apache developers using the FFM and found that there were three common types of personality profiles. Neuroticism and agreeableness were the two traits most important for differentiating the profiles from one another. They also found that FFM traits did not vary by developer role, membership, or degree of contribution to their project, and they demonstrated that developer personality was time-invariant. Further, they found a positive association between developers' openness to experience and the likelihood of making project contributions.

Finally, Iyer \emph{et al.}~\cite{Iyer.2021} examined the acceptance of pull requests in open-source projects. They found that pull requests from authors who score higher in openness and conscientiousness were more likely to be approved. They further found that extraversion was negatively related to the chance of approval. In addition, they found that pull requests which are closed by programmers who score higher on conscientiousness and extraversion had a greater chance of acceptance. Emotional stability of the closer, in contrast, was linked to a reduced likelihood of acceptance.

In the following, we will link each FFM personality trait in software developers to induced TD. First, extraversion indicates the degree of engagement with the external world. Sub-facets of extraversion include friendliness, gregariousness, assertiveness, activity, and excitement-seeking. Extraverted persons can thus be described as outgoing, often feeling positive emotions, and seeking stimulating activities. In comparison, introverted persons can be described as less outgoing, shy, and preferring to spend time alone~\cite{Gosling.2003, Matthews.2003, McCrae.1992}. Since more extraverted developers are more prone to activity and excitement-seeking, we propose that they ``move fast and break things'', leaving behind more TD than introverted developers. Extant literature further suggests that extraversion is positively related to risk-taking~\cite{Nicholson.2005}. Given that prior qualitative software engineering research argued that ``having a higher risk appetite can influence decisions to create technical debt''~\cite[p. 1504]{Tom.2013}, this further indicates that extraversion and TD might be positively connected. Formally put:

\emph{Hypothesis 1 (H1): Extraversion will be positively associated with induced TD.}

Agreeableness refers to the degree of concern with cooperation and social harmony. Sub-facets of agreeableness include trust, morality, altruism, cooperation, modesty, and sympathy. Agreeable persons can be described as friendly, helpful, and understanding. In comparison, non-agreeable persons can be described as less friendly, not very cooperative, and initiating disagreements~\cite{Gosling.2003, Matthews.2003, McCrae.1992}. Therefore, we expect more agreeable developers to wish to support others in improving code, thus paying down TD instead of adding to it. The literature on risk-taking also suggests that agreeableness is linked to less risk-taking~\cite{Nicholson.2005}. We therefore conjecture:

\emph{Hypothesis 2 (H2): Agreeableness will be negatively associated with induced TD.}

Conscientiousness indicates the degree of control over human impulses. Sub-facets of conscientiousness include orderliness, reliability, achievement-striving, and self-discipline. Thus, conscientious people can be described as prudent, organized, and reliable. In comparison, non-conscientious people can be described as impulsive, unsystematic, and less reliable~\cite{Gosling.2003, Matthews.2003, McCrae.1992}. Consequently, we expect more conscientious developers to dislike TD, and to induce little of it, or even remove it. The psychological literature on risk-taking takes a similar stance and finds an overall negative relationship between conscientiousness and risk-taking~\cite{Nicholson.2005}. We posit:

\emph{Hypothesis 3 (H3): Conscientiousness will be negatively associated with induced TD.}

Emotional stability is often described through its inverse, neuroticism. Neuroticism indicates the degree to which a person experiences negative feelings. Sub-facets of neuroticism include anxiety, anger, depression, self-consciousness, and vulnerability. Neurotic people can be described as tense, often in a bad mood, and emotional. In comparison, emotionally stable people can be described as calm and free from a persistent bad mood~\cite{Gosling.2003, Matthews.2003, McCrae.1992}. 
Neurotic individuals tend to act decisively to remedy the anxiety they are prone to experiencing in uncertain situations~\cite{Judge.2013} and to minimize potential threats that may materialize, even if this means accepting some concrete negative outcomes~\cite{Hirsh.2008}. In other words, they frequently resort to a ``better safe than sorry strategy''~\cite[p. 1005]{Lommen.2010} in uncertain situations. As discussed above, choices on TD are always made under uncertainty and involve trading off today's gains for potential future liabilities~\cite{Kruchten.2013b}, suggesting that neuroticism may lead to choices that avoid uncertainty by avoiding TD.
The psychological literature further discusses a potential link between neuroticism and self-oriented perfectionism~\cite{Flett.1989}, which may lead programmers to write whatever they perceive to be ``better'' code, which may well be code with less TD.
Again, the literature on risk-taking concurs with this idea and shows positive links between emotional stability and risk-taking~\cite{Nicholson.2005}. We thus hypothesize:

\emph{Hypothesis 4 (H4): Emotional stability will be positively associated with induced TD.}

Openness to experience describes the degree to which individuals are imaginative and creative. Sub-facets of openness include imagination, artistic interest, adventurousness, liberalism, and intellect. Persons high in openness can be described as individualistic, non-conforming, and aware of their feelings. In contrast, persons low in openness can be described as down-to-earth, conventional, and less aware of their feelings~\cite{Gosling.2003, Matthews.2003, McCrae.1992}. Consequently, we propose that developers higher in openness to experience will feel less bound to ideas of writing supposedly ``clean'' code, and are thus more likely to induce TD. Furthermore, prior literature has linked openness to experience with increased risk-taking~\cite{Nicholson.2005}. We thus propose:

\emph{Hypothesis 5 (H5): Openness will be positively associated with induced TD.}

\subsection{Regulatory Focus}

Despite the comprehensiveness of the FFM, it is common in the psychology literature to study it in conjunction with other traits~\cite{Marshall.2015}. We thus also study regulatory focus, an established construct in personality psychology that has not yet found any attention in software engineering research.
A person’s regulatory focus consists of two independent self-regulatory orientations, or foci, that shape their goal-striving behavior: Prevention focus and promotion focus~\cite{Higgins.1997, Lanaj.2012}.\footnote{Note that regulatory focus can be conceptualized as a state and a trait. As is also often done in the psychological literature, we only focus on the latter component, sometimes also referred to as ``chronic'' regulatory focus~\cite{Higgins.1997}.}
Promotion focus is related to being eager, risky, and oriented towards attaining gains as positive outcomes. In contrast, prevention focus is related to being careful, cautious, and oriented toward avoiding losses as negative outcomes. Given that incurring TD constitutes risk-taking, and in line with literature that links regulatory focus to risk-taking~\cite{Hamstra.2011}, we thus propose that more  promotion-focused developers may be more likely to induce TD, and more prevention-focused developers less so.

\emph{Hypothesis 6 (H6): Promotion focus will be positively associated with induced TD.}

\emph{Hypothesis 7 (H7): Prevention focus will be negatively associated with induced TD.}

\subsection{Narcissism}

Although there is an ongoing debate about the precise definition of the construct~\cite{Donnellan.2021}, scholars broadly agree that narcissism is a personality trait that combines ``a grandiose yet fragile sense of self and entitlement as well as a preoccupation with success and demands for admiration''~\cite[pp. 440--441]{Ames.2006}.
While various blog posts about narcissistic software developers suggest that narcissism may be a topic of interest in software engineering practice, we are not aware of any extant scientific work that explicitly studies this personality trait in software engineers.
Prior research in psychology, in contrast, is plentiful and has found that narcissism is related to a host of correlates, for example status-seeking~\cite{ZeiglerHill.2018}, having little regard for others' concerns, and viewing others as inferior~\cite{Morf.2001}.

A particularly prominent correlate of narcissism is risk-taking. Specifically, researchers have found associations of narcissism with a host of risky activities. These include, for instance, risky sexual behavior and sexual aggression, aggressive driving, drug and alcohol use, compulsive exercise, and gambling~\cite{Buelow.2018}. Given the nature of incurring technical debt as inherently risk-taking, we propose that more narcissistic developers are likely to induce more TD in their commits.

\emph{Hypothesis 8 (H8): Narcissism will be positively associated with induced TD.}

\section{Method}

\subsection{Technical Debt Measures}

To measure technical debt, we relied on the ``Technical Debt Dataset'' in version 2~\cite{Lenarduzzi.2019}. This dataset contains a comprehensive analysis of the master branches of 29 Apache open source projects using the popular tool SonarQube, and has been used in prior research~\cite{Codabux.2020}.

Our primary measure of induced TD in a focal commit follows prior work in relying on the estimated remediation effort for all maintainability issues as identified by SonarQube~\cite{Alfayez.2018}. Specifically, the value for the focal commit is the difference in remediation effort estimates between the focal commit and its parent commit. Positive values thus indicate increases in technical debt, and negative values indicate technical debt being paid down. In a secondary, broader measure of induced TD we additionally include the estimated remediation effort for all reliability and security issues as identified by SonarQube.

Because ratios are potentially problematic as dependent variables~\cite{Certo.2020}, we deviate from prior work~\cite{Alfayez.2018} and do not divide the TD measures by the difference in lines of code (LOC) between focal and parent commit. Instead, and arguably more precisely, we control separately for the LOC added and the LOC removed in the focal commit. This accounts for the fact that more LOC may make technical debt more likely~\cite{Alfayez.2018}.

\subsection{Survey Measures }

To obtain personality data, we surveyed the developers who made at least one commit in the Technical Debt Dataset. As the dataset itself does not contain contact information of developers, the first step was to obtain information on all commits of the projects from GitHub using PyDriller~\cite{Spadini.2018}. This yielded 99,972 commits, which partially extended beyond the timeframe of the dataset. We identified all individuals listed as ``authors'' in the data, and manually cleaned the list to remove duplicates. We also merged records where individuals used different names but the same email address, or different email addresses but the same or an extremely similar name for different commits. When judgment was needed, we made decisions as conservatively as possible. We ultimately obtained a list of 1,555 unique individuals. In case of multiple email addresses per person, we selected only one, preferring personal email addresses over professional ones because the person's commits to the projects were partially already quite old and the person might have moved organizations since. 

We sent an email to all 1,555 developers to invite them to our survey~\cite{Wagner.2020} (which was part of a larger data collection effort for multiple studies). We sent the invitation with a personalized link to the survey. In the email, we pledged to donate US\$ 2 per completed response to the United Nations World Food Programme~\cite{Baltes.2022}. We also sent two reminders~\cite{Wagner.2020}, including one that linked to an official university web page confirming the authenticity of the survey as some developers were concerned that the invitation might be a scam.

Because 165 emails bounced, we reached 1,390 developers (89.4\% deliverable emails), of which 194 developers started the survey, and 124 completed it. We dropped all respondents that provided implausible values for their age ($\leq$10 or $\geq$100), leaving us with a preliminary sample of 121 developers.
Although this number may seem low, it is of a similar magnitude as the total number even from studies that assessed personality not based on self-reports but on mining vast corpora of emails. Calefato \emph{et al.}, for instance, used about 1.35 million emails from 46,304 developers but only obtained full personality profiles for 211 of them ($<$0.5\%), of which only 118 had made source code commits~\cite{Calefato.2019}. Our response rate was 8.9\%, which is similar to that of other studies surveying developers on GitHub. Graziotin \emph{et al.}, for instance, report a 7\% response rate \cite{Graziotin.2017}. Their reported share of deliverable emails is 96.6\%, which is also similar to ours.

The first page of the survey provided participants with basic information about what kind of data would be collected. We assured the developers that their data would be treated completely confidentially and not be shared with other parties. We highlighted that participation was voluntary and that developers could abort the survey at any point. We further explicitly clarified that the developers consented to participating in our study by proceeding to the next page. We pretested the survey with three doctoral students of software engineering and made minor modifications based on their feedback.

To capture respondents' FFM personality traits, we employed the widely used Ten-Item Personality Measure (TIPI)~\cite{Gosling.2003}. Since the TIPI is designed to capture all facets of the Big Five with content and criterion validity with one item each, reliability measures like Cronbach's $\alpha$ are uninformative~\cite{Gosling.}. We consequently do not report them.

To measure regulatory focus, we employed six items (three each for promotion and prevention focus) from the extensively validated Regulatory Focus Composite Scale (RF-COMP)~\cite{Haws.2010}. To test for internal consistency, we performed a factor analysis on all ten items of the RF-COMP in our sample. It indicated more than the two expected distinct factors. As is commonly done in such cases, we therefore reduced the number of items until two clear factors emerged---one capturing promotion and one capturing prevention focus. We then obtained a Cronbach's $\alpha$ of .66 for promotion focus and .60 for prevention focus. Given the low number of items per construct, these constitute acceptable values~\cite{Cortina.1993}. 

We measured narcissism using the short version of the Narcissistic Personality Inventory (NPI-16)~\cite{Ames.2006}. Cronbach's $\alpha$ was .70, constituting an acceptable value~\cite{Nunnally.1978} and even being slightly higher than that obtained in the work that developed the measure in the first place~\cite{Ames.2006}.

As we measure personality after the analyzed commits were made, this might potentially raise concerns about an unclear direction of causality in our analysis. However, this is unlikely to be a problem as personality is widely considered as stable across adult life \cite{Costa.1988, McCrae.1994} and the sampled developers were all between 24 and 65 years of age. In line with this, recent research specifically on the personality of developers has not found any evidence of variation over time~\cite{Calefato.2019}.

To measure developers' age at the time of each commit, we asked developers to simply provide their age in years. We then subtracted the difference between 2022 and the year in which the focal commit was made from the provided age.

\subsection{Final Sample}

We only retained normal commits, dropping merge and orphan commits because the former do not allow a calculation of induced TD (due to multiple parent commits) and the latter may have particular characteristics. Note that including orphan commits does not change our results.

Since our TD measures require a completed SonarQube analysis for the focal and the parent commit, and there was substantial missing data in the Technical Debt Dataset, our sample was ultimately reduced to $N$ = 2,145 commits from 19 developers for which we had both full TD and personality data. While our sample is thus not large by any means, it is not too far from, for example, the sample of 28 individuals in the study of Calefato and Lanubile on personality and trust between Apache developers~\cite{Calefato.2018b} and much larger than Rigby and Hassan's sample of four programmers in their study on the FFM traits of highly productive Apache developers~\cite{Rigby.2007}.

\subsection{Analysis Strategy}

We ran panel regressions to study the association between our personality and TD measures. This method is suitable in situations in which a panel unit (the developer in our case) is observed repeatedly. We clustered standard errors at the panel unit to account for the fact that multiple commits from the same developer are not statistically independent. We included control variables for developer age at time of commit and LOC added and LOC removed in our models. In addition, we include dummy variables (fixed effects) for each project in our analyses. These variables capture all time-invariant aspects that are specific to a project, for example certain coding conventions.

\subsection{Replication Package}

Unfortunately, we cannot make the data for our analyses available to other researchers because we explicitly promised our respondents full confidentiality. Even if this were different, we would be hesitant to share the data because we could not do it anonymously given that the entire purpose of this paper is to link personality to individually identifiable contributions to software projects. The Technical Debt Dataset itself, however, is publicly available~\cite{Lenarduzzi.2019}. We also provide a package of all our analysis scripts.\footnote{Available at: https://dx.doi.org/10.6084/m9.figshare.21802968} This package also contains a dummy data file with the precise wording of the used survey measures.

\section{Results}

\subsection{Descriptive Statistics}

\begin{figure}[h]
  \centering
  \includegraphics[width=\linewidth]{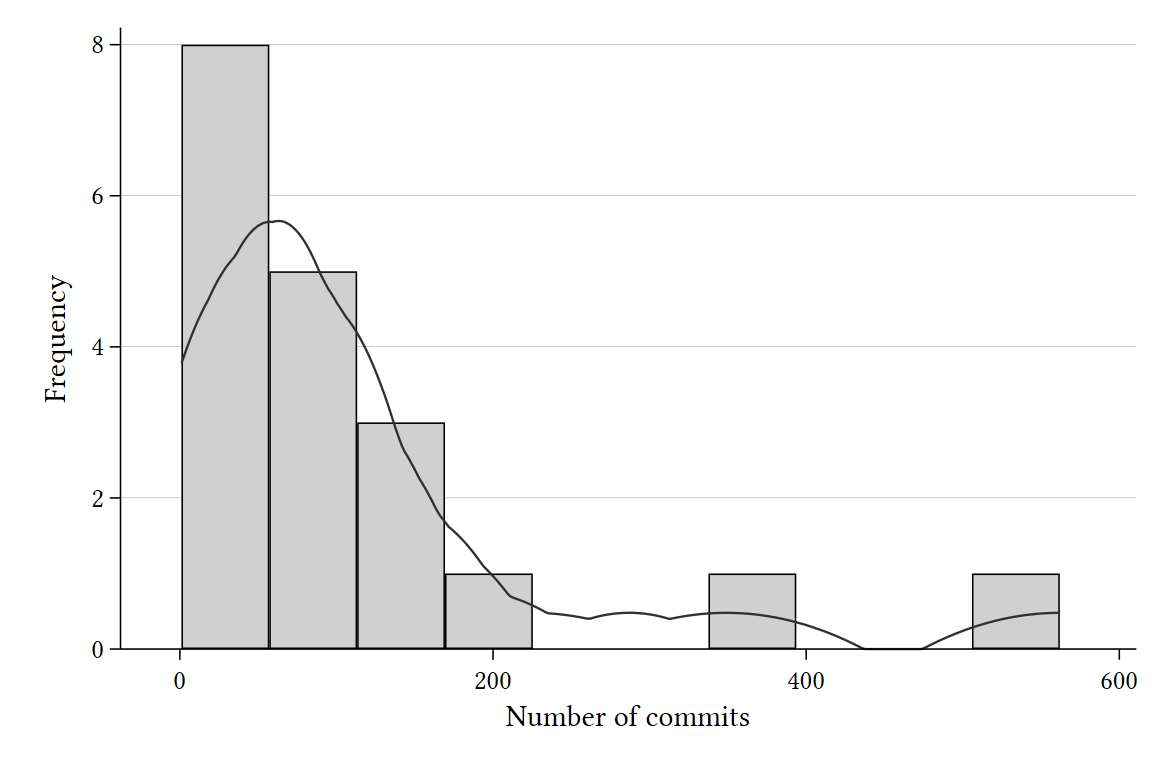}
  \caption{Histogram and Kernel Density Plot of Commits per Developer\label{fig:histogram_developer_commits}}
\end{figure}

Table \ref{tab:descriptives} provides descriptive statistics for the final dataset. Figure \ref{fig:histogram_developer_commits} shows a histogram and kernel density plot of the number of commits per developer in the sample.

\begin{table}[ht]
\centering
\def\sym#1{\ifmmode^{#1}\else\(^{#1}\)\fi}
\small
\caption{Descriptive Statistics \label{tab:descriptives}}
\begin{tabular}{lcccc} 
\toprule
                    &        Mean&          SD&         Min.&         Max.\\
\midrule

Induced TD          &       65.41&    1,059.15&  -14,160.00&   32,880.00\\
Induced TD (broad)  &       67.92&    1,091.94&  -14,320.00&   33,829.00\\
Extraversion        &        3.57&        1.36&        1.00&        6.00\\
Agreeableness       &        4.18&        1.10&        2.50&        6.00\\
Conscientiousness   &        5.82&        1.11&        3.50&        7.00\\
Emotional stability &        4.83&        1.06&        1.50&        6.50\\
\makecell[l]{Openness to \\ \thickspace experience}&        4.82&        1.25&        3.00&        7.00\\
Promotion focus     &       15.30&        2.39&       10.00&       21.00\\
Prevention focus    &       13.13&        3.52&        9.00&       18.00\\
Narcissism          &        3.66&        2.14&        0.00&        9.00\\
Age at commit       &       36.48&        6.88&       23.00&       50.00\\
LOC added           &      217.63&    1,565.86&        0.00&   40,180.00\\
LOC removed         &       81.12&      732.66&        0.00&   25,886.00\\

\bottomrule
\end{tabular}
\end{table}

Table \ref{tab:corr} shows the correlation between all variables except for the project dummy variables. It is apparent that the two measures of TD are almost perfectly correlated and thus tap into essentially the same construct.
Unsurprisingly, there is also a strong correlation between induced TD and LOC added.

\begin{table*}[ht]
\centering
\def\sym#1{\ifmmode^{#1}\else\(^{#1}\)\fi}
\footnotesize
\caption{Correlations\label{tab:corr}}
\begin{tabular}
{lcccccccccccc}
\toprule

                &\makecell{Induced\\TD}         &\makecell{Induced\\TD\\(broad)}         &\makecell{Extra-\\version}         &\makecell{Agree-\\able-\\ness}         &\makecell{Consci-\\entious-\\ness}         &\makecell{Emo-\\tional\\stabi-\\lity}         &\makecell{Open-\\ness to\\expe-\\rience}         &\makecell{Pro-\\motion\\focus}         &\makecell{Pre-\\vention\\focus}         &\makecell{Narcis-\\sism}         &\makecell{Age\\at\\commit}         &      \makecell{LOC\\added} \\

\midrule
\makecell[l]{Induced TD\\ \thickspace (broad)}&     1.00\sym{***}&              &                  &                  &                  &                  &                  &                  &                  &                  &                  &                                    \\
\makecell[l]{Extraversion}    &     0.02         &     0.02         &              &                  &                  &                  &                  &                  &                  &                  &                  &                                    \\
\makecell[l]{Agreeableness}   &     0.02         &     0.02         &    -0.30\sym{***}&              &                  &                  &                  &                  &                  &                  &                  &                                    \\
\makecell[l]{Conscien-\\ \thickspace tiousness}&    -0.04\sym{*}  &    -0.04\sym{*}  &    -0.27\sym{***}&    -0.55\sym{***}&              &                  &                  &                  &                  &                  &                  &                                    \\
\makecell[l]{Emotional \\ \thickspace stability}&    -0.02         &    -0.02         &    -0.31\sym{***}&     0.12\sym{***}&     0.26\sym{***}&              &                  &                  &                  &                  &                  &                                    \\
\makecell[l]{Openness to \\ \thickspace experience}&     0.04         &     0.04         &     0.42\sym{***}&     0.13\sym{***}&    -0.41\sym{***}&    -0.58\sym{***}&              &                  &                  &                  &                  &                                    \\
\makecell[l]{Promotion \\ \thickspace focus} &    -0.04\sym{*}  &    -0.04\sym{*}  &    -0.20\sym{***}&     0.29\sym{***}&    -0.03         &    -0.13\sym{***}&    -0.15\sym{***}&              &                  &                  &                  &                                    \\
\makecell[l]{Prevention \\ \thickspace focus}&    -0.02         &    -0.02         &     0.15\sym{***}&     0.30\sym{***}&    -0.43\sym{***}&    -0.72\sym{***}&     0.46\sym{***}&     0.57\sym{***}&              &                  &                  &                                    \\
\makecell[l]{Narcissism}      &    -0.02         &    -0.02         &     0.45\sym{***}&    -0.30\sym{***}&     0.00         &    -0.00         &    -0.06\sym{**} &    -0.46\sym{***}&    -0.18\sym{***}&              &                  &                                    \\
\makecell[l]{Age at commit}   &     0.02         &     0.02         &    -0.46\sym{***}&     0.39\sym{***}&    -0.33\sym{***}&     0.20\sym{***}&    -0.41\sym{***}&     0.02         &    -0.11\sym{***}&    -0.19\sym{***}&              &                                    \\
\makecell[l]{LOC added}       &     0.73\sym{***}&     0.73\sym{***}&    -0.01         &     0.04         &    -0.05\sym{*}  &    -0.02         &     0.07\sym{***}&    -0.05\sym{*}  &    -0.01         &    -0.04\sym{*}  &     0.04         &                                \\
\makecell[l]{LOC removed}     &    -0.13\sym{***}&    -0.13\sym{***}&    -0.01         &     0.03         &    -0.03         &     0.01         &     0.03         &    -0.01         &    -0.01         &    -0.01         &    -0.01         &     0.39\sym{***}              \\

\bottomrule

\multicolumn{13}{l}{\footnotesize \sym{*} \textit{p} $<$ 0.05, \sym{**} \textit{p} $<$ 0.01, \sym{***} \textit{p} $<$ 0.001. Dummy variables for projects not shown. $N$ = 2,145. }
\end{tabular}
\end{table*}

It is well-documented and thus not surprising that FFM traits and narcissism share correlations~\cite{Visser.2018}. In our sample we replicate, for instance, the positive correlation between narcissism and extraversion and the negative correlation between narcissism and agreeableness that have been widely reported in the psychological literature~\cite{Visser.2018}. As we expected from regulatory focus measurements in other populations, promotion and prevention focus are correlated positively and significantly with each other~\cite{Lanaj.2012}.

\subsection{Findings}

Table \ref{tab:regressions} reports the regression coefficients and standard errors of our panel regression analyses, once with the primary measure of induced TD and once with the secondary, broader measure of induced TD as the dependent variable. As is evident from the table, the results are similar across both dependent variables. Specifically, the coefficients for extraversion and agreeableness are negative and not significant. These results thus do not lend support to \emph{H1} and \emph{H2}. The coefficient for conscientiousness is negative and significant, supporting \emph{H3}. Counter to our hypotheses, the coefficients for both emotional stability and openness to experience are significant but negative. These findings consequently constitute evidence against \emph{H4} and \emph{H5}. The results regarding regulatory focus are mixed. The coefficient for promotion focus is unexpectedly negative but not significant. The coefficient for prevention focus is negative as expected, and significant. These results thus do not support \emph{H6} but corroborate \emph{H7}. Finally, the coefficient for narcissism is positive but not significant, lending no support to \emph{H8}.

\begin{table}[ht]
\centering
\def\sym#1{\ifmmode^{#1}\else\(^{#1}\)\fi}
\small
\caption{Panel Regressions \label{tab:regressions}}
\begin{tabular}{lcc} 
\toprule

 &\makecell{Induced \\TD} & \makecell{Induced\\ TD (broad)} \\

\midrule
Extraversion        &      -34.80         &      -36.55         \\
                    &     (45.98)         &     (48.32)         \\
\addlinespace
Agreeableness       &       -0.93         &       -1.91         \\
                    &     (25.17)         &     (26.32)         \\
\addlinespace
Conscientiousness   &     -185.20\sym{**} &     -191.80\sym{**} \\
                    &     (60.77)         &     (63.85)         \\
\addlinespace
Emotional stability &     -178.47\sym{***}&     -183.47\sym{***}\\
                    &     (21.78)         &     (22.74)         \\
\addlinespace
Openness to experience&     -189.82\sym{***}&     -192.61\sym{***}\\
                    &     (43.50)         &     (45.14)         \\
\addlinespace
Promotion focus     &      -31.66         &      -31.41         \\
                    &     (28.94)         &     (30.36)         \\
\addlinespace
Prevention focus    &      -44.31\sym{***}&      -45.70\sym{***}\\
                    &      (7.09)         &      (7.44)         \\
\addlinespace
Narcissism          &       19.77         &       22.23         \\
                    &     (20.67)         &     (21.52)         \\
\addlinespace
Age at commit       &      -17.77\sym{*}  &      -17.94\sym{*}  \\
                    &      (7.51)         &      (7.78)         \\
\addlinespace
LOC added           &        0.62\sym{***}&        0.64\sym{***}\\
                    &      (0.14)         &      (0.14)         \\
\addlinespace
LOC removed         &       -0.71\sym{***}&       -0.73\sym{***}\\
                    &      (0.14)         &      (0.14)         \\
\addlinespace
Constant            &     4500.48\sym{***}&     4596.41\sym{***}\\
                    &   (1152.82)         &   (1203.54)         \\
\midrule
Project fixed effects        &        Yes         &        Yes         \\
Observations        &        2,145         &        2,145         \\
Clusters              &        19             &       19              \\

\bottomrule
\multicolumn{3}{l}{\footnotesize Dependent variable indicated in top row.}\\
\multicolumn{3}{l}{\footnotesize Table reports coefficients, clustered standard errors in parentheses.}\\
\multicolumn{3}{l}{\footnotesize \sym{*} \textit{p} $<$ 0.05, \sym{**} \textit{p} $<$ 0.01, \sym{***} \textit{p} $<$ 0.001}\\
\end{tabular}
\end{table}

While there is substantial disagreement in the literature regarding whether corrections for multiple testing are necessary and whether and when they help the interpretation of results~\cite{Rothman.1990, Bender.2001, OKeefe.2003} (especially in the context of multiple regression), we attempted to control the false discovery rate by computing adjusted \textit{p}-values (sharpened \textit{q}-values)~\cite{Anderson.2008} 
for all eight independent variables of interest. The results reaffirm our original findings.

The coefficients of our unhypothesized control variables are significant and with unsurprising signs. Developer age at the time of the commit was negatively associated with induced TD. LOC added and LOC removed were positively and negatively associated with induced TD, respectively.

\section{Discussion}

\subsection{Unexpected results}

While some hypotheses were simply not supported by insignificant results in our regression analyses, some significant results went directly counter to our hypotheses and therefore warrant further discussion.
First, emotional stability was significantly negatively related to TD. A possible explanation---in line with psychological research linking positive affect to enhanced problem solving and decision making, as well as flexible, innovative, thorough, and efficient cognitive processing~\cite{Isen.2001}---would be that more emotionally stable developers' generally positive mood allows them to find superior ways to produce highly maintainable code, inducing less TD.

Second, openness to experience was negatively related to TD. 
One plausible  explanation would be that individuals high in openness possess a specific aesthetic sensitivity~\cite{John.2021b}, which might find an expression in the creation of particularly beautiful code, which in turn may include little TD. An alternative explanation would be that such individuals are highly creative~\cite{John.2021b} and may thus simply find good ways of expressing logic in code without creating TD in the first place.

\subsection{Threats to validity}

There are several potential issues that may threaten the validity of our study. We treat them along four categories~\cite{Gren.2018}.

\subsubsection{Construct validity}

One may challenge the accuracy of our measurements. This holds true for both the independent and the dependent variables. When it comes to the personality measures, we resorted to using short scales. While this is likely to have contributed to a higher response rate in our survey, these measures may have introduced more measurement error than longer scales would have~\cite{Schmidt.1996}.
Additionally, our personality measures are based on self-reports. Although such measures are the most trusted and most frequently used in psychology, they are potentially not problem-free~\cite{Paulhus.2007}. For instance, respondents may seek to present themselves in the best possible light rather than wholly accurately.
Combined with limited response rates among software engineer populations, this has made researchers look to other data sources. For instance, it has recently become fashionable to infer personality from text corpora \cite{Calefato.2019, Rastogi.2016, ParumaPabon.2016, Iyer.2021}, but such methods are at least as problematic as self reports. For one, recent work demonstrated that using psycholinguistic methods to capture personality from developer communication (e.g., mailing lists) can result in substantial inaccuracies~\cite{vanMil.2021} and different linguistic tools may lead researchers to very different conclusions~\cite{Calefato.2022}. For another, most such work has only studied the FFM personality traits and corresponding linguistic methods are not readily available for other personality traits.

Despite the fact that ``few findings in psychology are more robust than the stability of personality'' \cite[p. 175]{McCrae.1994}, this notion has been challenged in some software engineering research, suggesting that contributors evolve as more conscientious, more extrovert and less agreeable over the years of participation \cite{Rastogi.2016}. While the effect sizes in this research are tiny\cite{Calefato.2019}, this potentially affects the validity of our personality measures.

Further, the TD estimates from SonarQube are not perfect and can likely be improved upon~\cite{Lenarduzzi.2019b}. In fact, there are likely limits to automated detection of TD in general. Certainly, source code analysis tools cannot capture all types of technical debt that prior research has identified~\cite{Tom.2013, Alves.2016, Rios.2018b}. Relatedly, different technical debt detection tools frequently come to different assessments of technical debt~\cite{Lefever.2021}. It would thus be interesting to repeat our analysis with different measures of technical debt from other tools.

\subsubsection{Internal validity}

There are various ways in which the internal validity of our study might be limited. First, developer personality may not be the only driver of TD, and not all other relevant variables might be controlled for. This could lead to missing variable problems and hence to spurious findings in our regression analyses.

Second, our results might be affected by selection effects. 
Since not all authors in the studied software projects have commit privileges, the commits that were made and that we can consequently observe may differ systematically from all proposed code changes. For instance, as others have shown, the personalities of committers influence their behavior, too~\cite{ParumaPabon.2016, Iyer.2021}. Consequently, our data may be filtered in particular ways, and our analyses and findings might thus be distorted.

\subsubsection{External validity}

Naturally, a key question in software engineering studies on data from software repositories relates to how generalizable any findings are to other contexts.
At least two potential threats to validity exist in this regard.

First, in light of recent discussions about the representativeness of samples \cite{Baltes.2022, Wagner.2020}, we highlight that our sample is likely not representative of all software developers. In particular, we only sample open source contributors from relatively large Apache projects written largely in Java, and we achieved only a limited response rate in our survey. 

Second, the Technical Debt Dataset exhibited substantial missing data. In particular, the SonarQube analyses were incomplete for many 
commits. Since our measure of TD requires completed analyses for not only each focal commit but also its parent commit, we experienced a lot of missing values in our dependent variables. Whether this data is missing at random remains unclear.

\subsubsection{Reliability}
The reliability of our research should be high. All scales we used to assess personality are established in the psychological literature~\cite{Wagner.2020}. There was very little human judgment needed in our research process. Where judgment calls were necessary, we documented the guiding principles in this article. We also provide all scripts used for data analysis. Consequently, our research should be, in principle, fully replicable. Given that confidentiality assurances to the surveyed developers prevent us from sharing the personality data, other researchers are of course not able to directly repeat our analyses. However, we deem the latter issue not a threat to the validity of our research.

\subsection{Implications for Research}

Our study has several implications for future research.
First and foremost, it demonstrates that developer personality has a meaningful impact on TD. We thus propose further study in this direction. In particular, we encourage replications of our study in other code bases and potentially with other TD measures, e.g., self-admitted technical debt. Further, studying other types of TD that may not be captured by the measures we used may be informative~\cite{Tom.2013, Alves.2016, Rios.2018b}. Moreover, potential consequences of TD like software defects or project budget overruns might also be linked to personality and warrant further study.
Second, our research shows that personality traits beyond the FFM may be relevant to explain and predict the behavior of software developers. Although only prevention focus proved to be significant in our study, we suggest that regulatory focus in general and possibly even narcissism should be further studied with regard to software engineering decisions beyond the issue of TD.
Finally, future research may consider the addition and the removal of TD as two distinct activities. We thus propose analyses of commits with net increases and commits with net decreases in TD~\cite{Alfayez.2018} to see if they may potentially have different personality antecedents.

\subsection{Implications for Practice}

Our study suggests that developers with certain personality profiles might be more or less prone to incurring TD. This leads to important practical implications. First, project managers (who often know their teams well and do not need to administer potentially legally problematic personality surveys) might staff software development projects with this in mind. For instance, they might purposefully opt to assign project artifacts that already have too much TD
to developers with a personality profile that likely leads to reduced TD, for example, developers high in prevention focus or conscientiousness. Similarly, they may change team composition between project phases. Following Beck's 3X model, it is for example conceivable that the ``explore'' phase allows for developers producing more TD, whereas ``expand'' calls for more careful developers, and the ``extract'' phase may be best handled by developers averse to TD~\cite{Beck.2016}.
Second, managers may allocate more time and resources to closely tracking TD in artifacts developed by programmers with a personality profile that is conducive to TD.
Third, when using techniques like pair programming, project managers may wish to make personality a criterion in establishing pairs. For example, it may be desirable to have pairs of programmers that have complementary personality profiles to actively manage TD in a project.
Finally, our research may allow developers themselves to become more aware of their innate tendencies. All personality measures used in this study are generally available and could be used in training programs in which developers have an opportunity to assess their personality and learn about the potential implications for their behaviors regarding TD.

\section{Conclusion}

In conclusion, it seems that developer personality influences the creation and removal of TD. We hope our research sparks further inquiries into this intriguing issue.

\section*{Acknowledgment}

We would like to acknowledge helpful information on the Technical Debt Dataset from Davide Taibi.

\bibliographystyle{IEEEtran}
\bibliography{IEEEabrv,base.bib}

\end{document}